\documentclass[a4paper,10pt]{article}
\pdfoutput=1

\usepackage{hyperref}

\usepackage{
graphicx,
hyperref,
amsmath,
amssymb,
charter,
xcolor,
ifluatex,
multirow}
        
\definecolor{Gray}{gray}{0.95}
\definecolor{RGray}{gray}{0.85}
\definecolor{CGray}{gray}{0.92}

\definecolor{tit}{rgb}{0.1,0.2,0.4}
\definecolor{blus}{cmyk}{1,1,0,0.6}
\definecolor{verde}{cmyk}{0.92,0,0.59,0.25}

\usepackage{tipa}
\usepackage{amsmath,amssymb,amsfonts, bm}
\usepackage{epsfig}
\usepackage{graphicx}
\usepackage{slashed}
\usepackage{color}

\usepackage{caption}
\usepackage{subcaption}
\usepackage{slashed} 
\usepackage{float}
\usepackage{comment}

\usepackage{booktabs}
\usepackage[normalem]{ulem}
\newcommand{\eps}{\epsilon}

\newcommand{\D}{{\cal D}}
\newcommand{\U}{{\cal U}}

\newcommand{\N}{{\cal N}}
\newcommand{\M}{{\cal M}}

\usepackage{calc}

\newcommand{\be}{\begin{equation}}
\newcommand{\ee}{\end{equation}}

\newcommand{\bea}{\begin{eqnarray}}
\newcommand{\eea}{\end{eqnarray}}

\newcommand{\bfig}{\begin{figure}}
\newcommand{\efig}{\end{figure}}

\makeatletter
\newcommand*{\rom}[1]{\expandafter\@slowromancap\romannumeral #1@}
\makeatother

\usepackage{tikz}

\usepackage[margin=1in]{geometry}

\usetikzlibrary{arrows.meta, positioning}

\usetikzlibrary{arrows.meta, positioning, decorations.pathmorphing}

\begin{document}
\allowdisplaybreaks
\vspace*{-2.5cm}
\begin{flushright}
{\small
IIT-BHU
}
\end{flushright}

\vspace{2cm}

\begin{center}
{\LARGE \bf \color{tit} Return of the technicolour  }\\[1cm]

{\large\bf Gauhar Abbas$^{a}$\footnote{email: gauhar.phy@iitbhu.ac.in}\\  [7mm]
}

\vspace{1cm}
{\large\bf\color{blus} Abstract}
\begin{quote}

\end{quote}

\thispagestyle{empty}
\end{center}

\begin{quote}
{\large\noindent\color{blus} 
}
We discuss that conventional Technicolour dynamics can be revitalized within the Dark Technicolour  paradigm by invoking the Extended Most Attractive Channel  hypothesis. In this framework, Standard Model  fermions acquire masses via multifermion chiral condensates arising from new strong dynamics. The model incorporates three confining gauge sectors, Technicolour, Dark Technicolour, and an intermediate QCD-like sector, linked through extended gauge symmetries. The  Extended Most Attractive Channel  hypothesis reveals a hierarchical structure of condensates, where channels with higher net chirality become increasingly attractive. At low energies, the Dark Technicolour  paradigm naturally reduces to the Froggatt–Nielsen or Standard Hierarchical Vacuum Expectation Value  model, governed by residual discrete symmetries, offering a compelling resolution to the Standard Model Flavor Problem.
\end{quote}

\vspace{8cm}

Based on the talk given in APGC-2025. 

\newpage
\setcounter{footnote}{0}
\section{Introduction}
At the most fundamental level, our Universe is profoundly symmetric. This fact is an evidence of symmetries our Universe was born with.  This elemental disposition is manifested even in the classical realm.  One can observe this in the astounding patterns found throughout in nature, for example, in flowers, plants, and birds.  At first glance, these patterns apparently appear highly symmetrical.  However, a closer examination reveals that symmetries are subtly broken, leading to a deeper complexity beneath the surface.

%

Some of the symmetries of our Universe are now well understood.  For instance, the visible ordinary matter, which comprises of approximately  $5\%$  of the Universe, is governed by the symmetry group $ \rm SU(3)_c \times SU(2)_L \times U(1)_Y$ of the Standard Model (SM) of Particle Physics.  The SM is the most successful quantum theory capable of  describing the known constituents and interactions of our Universe. The truly remarkable fact is that the SM reveals an amazing and subtle pattern of symmetry even at a superficial level.  For example, there are three fermionic families and three leptonic families. However, an intriguing observation is the numerical relation:
\begin{align}
    \text{Number of fermions} &=  \text{Number of gauge bosons} \\ \nonumber
   & =  \text{Number of symmetry generators  of the SM}  =12.
\end{align}

There may still exist  fundamental symmetries of the Universe that are  unknown to us. This possibility is underscored by the fact that the SM symmetry group $ \rm SU(3)_c \times SU(2)_L \times U(1)_Y$ accounts  for only about  $5\%$  of total content of the Universe.  The remaining content consists of dark-matter, which constitutes  of  approximately  $27\%$,  and dark energy, which  makes up the rest.

This raises a compelling question: if
\begin{align}
 5\% \ \text{of the Universe} \in \rm SU(3)_c \times SU(2)_L \times U(1)_Y,
\end{align}
then perhaps
\begin{align}
 27\% \ \text{of the Universe} \in \text{a larger symmetry group, such as} \ \rm SU(N) \times SU(N) \times \cdots \times U(1) \times \cdots \times U(1).
\end{align}

\begin{figure}[h]
	\centering
 \includegraphics[width=0.7\linewidth]{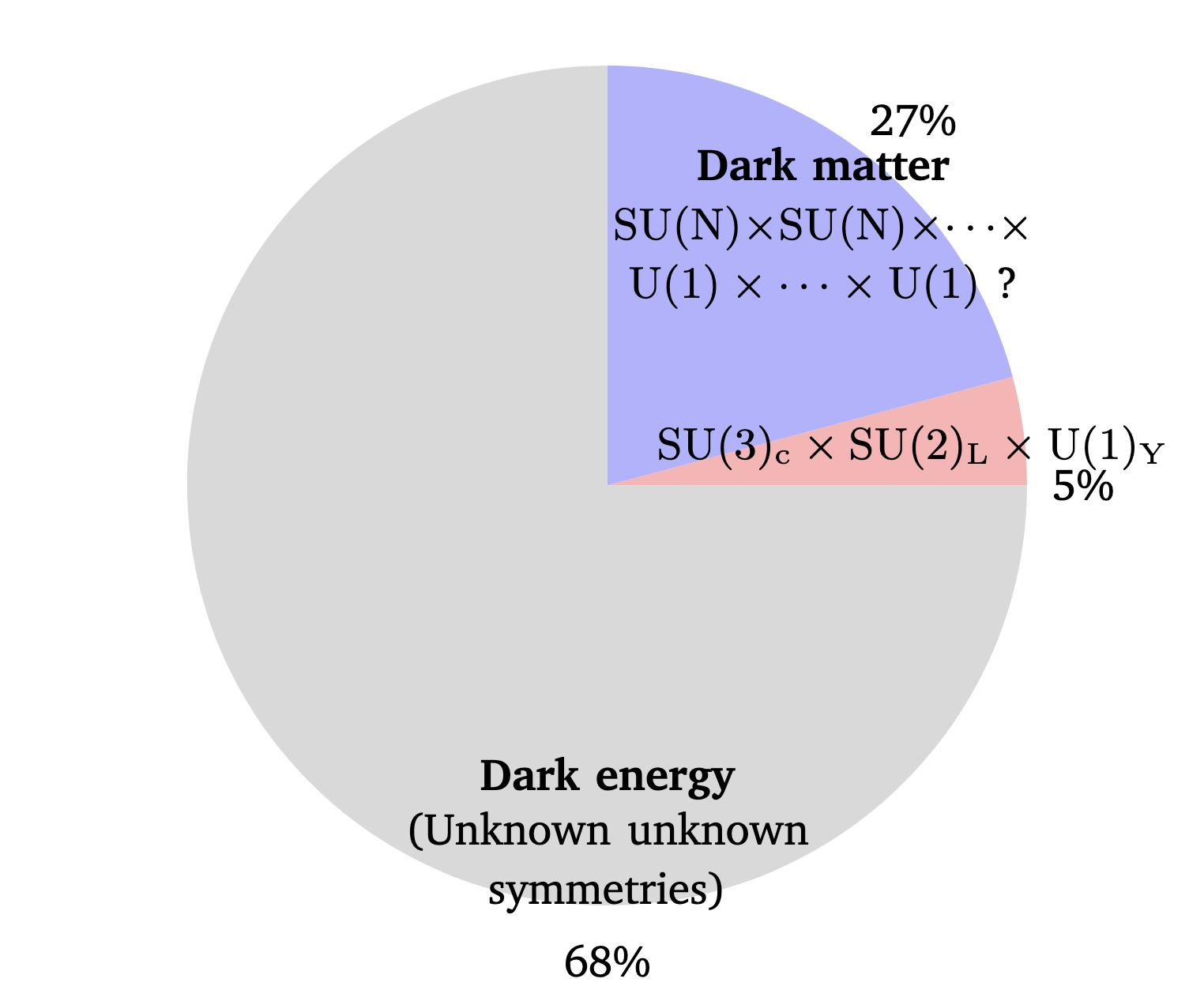}
 \caption{ The  matter problem }
  \label{fig2}
	\end{figure}

This perspective suggests, as depicted in figure \ref{fig2},  that dark matter may be governed by an extended or hidden gauge structure, hinting at a richer symmetry landscape beyond the SM (BSM).
    
A central question in modern physics is: where does the mass of the Universe's constituents come from? For ordinary visible matter, the answer lies in the Higgs mechanism. However, the introduction of the Higgs field into the SM is somewhat ad hoc and leads to several unresolved theoretical issues.

To understand this, recall that the electroweak symmetry group $\rm SU(2)_L \times U(1)_Y$ is spontaneously broken to the electromagnetic subgroup $\rm U(1)_{EM}$ by the vacuum expectation value (VEV) of the Higgs field. This symmetry-breaking process, known as the Higgs mechanism, is encoded in the SM Lagrangian:
\begin{align}
\mathcal{L}_{\rm SM} = \mathcal{L}_{\rm gauge, fermions} + \mathcal{L}_{\rm Higgs, gauge, fermions}.
\end{align}

The first term, involving gauge fields and fermions, is highly symmetric and remains stable under quantum corrections. In contrast, the second term, which includes the Higgs field responsible for generating the masses of fermions and gauge bosons, is extremely sensitive to quantum corrections, making it one of the most theoretically fragile components of the SM.

This sensitivity becomes clear when we examine the form of the Higgs potential:
\begin{align}
V(\varphi) = -\mu^2 \varphi^\dagger \varphi + \lambda (\varphi^\dagger \varphi)^2.
\end{align}
The spontaneous symmetry breaking (SSB) is triggered by the negative mass-squared term ($-\mu^2$).  However, the stability of the negative mass-squared term is marred by large quantum corrections:
\begin{align}
  \Delta \mu^2 \sim \Delta m_h^2 \sim \Lambda^2,
\end{align}
where $\Lambda$ is the cut-off scale of the SM.  This defines the well-known hierarchy problem: what stabilizes the Higgs boson mass near the electroweak scale in the presence of quadratically divergent corrections?

Over the past several decades, a significant effort has been devoted to constructing ``Super Beautiful and Incredible" (SUBI) theories that elegantly solve the hierarchy problem. These frameworks predicted new physics near the electroweak scale. However, despite extensive searches at the LHC up to several TeV, no such new particles  have been found.

In essence, the LHC has revealed only two critical facts:
\begin{enumerate}
\item The Higgs boson exists, with a measured mass of 125.20 GeV.
\item A hierarchy exists between the Higgs mass and the scale of any potential new physics.
\end{enumerate}

The absence of new weak-scale states at the LHC, together with the measured Higgs mass of 125.20 GeV, may indicate that the conventional formulation of the hierarchy problem is incomplete. In particular, the quadratic sensitivity of the Higgs mass to ultraviolet scales is derived under the assumption that the Higgs field is a fundamental degree of freedom up to arbitrarily high energies.   Thus, the hierarchy problem may be a symptom of an incorrect ultraviolet (UV) assumption, namely, that the Higgs is fundamental.  It may be time to reassess whether the hierarchy problem should  remain the central guiding principle for BSM model building. Instead, future directions may be better motivated by observational imperatives such as the nature of dark matter, the origin of the matter–antimatter asymmetry,  and the flavour structure of the SM.

A pressing and meaningful question then emerges: can the stability of the Higgs mass be achieved naturally within a framework that also accounts for these profound mysteries?  Can we have a ``Sweet and Inteligent" (SWEETI) theory capable of explaining, for example, the flavour structure of the SM (the so-called flavour problem)~\cite{Abbas:2023ivi,Abbas:2024wzp},  and at the same time stabilizing the Higgs mass with a hierarchical spectrum? From this point forward, we remain focused on addressing this question.

An alternative perspective is that the Higgs boson may instead be a low-energy composite state, emerging from new strong dynamics at the TeV scale. In such a scenario, the Higgs mass is controlled primarily by the dynamics of confinement and chiral symmetry breaking, rather than by UV physics. The apparent hierarchy between the electroweak scale and higher scales then reflects the separation between emergent and fundamental degrees of freedom, rather than an instability requiring fine-tuning. 

From this viewpoint, the hierarchy problem is not eliminated, but reinterpreted: it signals the breakdown of the SM description above the compositeness scale. This shift in perspective motivates frameworks in which electroweak symmetry breaking, fermion masses, and flavour hierarchies arise dynamically from strongly coupled sectors. The dark-technicolour paradigm explored in this work is a concrete realization of this idea.

We begin by defining the flavour problem of the SM. Interestingly, the flavour structure of the SM is tightly linked to the Higgs field and is encoded in the Yukawa Lagrangian:
\bea
\label{mass1}
-{\mathcal{L}}_{\rm Yukawa} &=&     y^u_{ij} \bar{ \psi}_{L_i}^q  \tilde{\varphi} \psi_{R_j}^{u}
+ y^d_{ij} \bar{ \psi}_{L_i}^q  \varphi \psi_{R_j}^{d}
+ y^\ell_{ij} \bar{ \psi}_{L_i}^\ell  \varphi \psi_{R_j}^{\ell}   + \text{H.c.}.  \nonumber
\eea
    where $L$ and $R$ denote left- and right-handed charged-fermion fields, respectively.

\begin{figure}[H]
	\centering
 \includegraphics[width=0.9\linewidth]{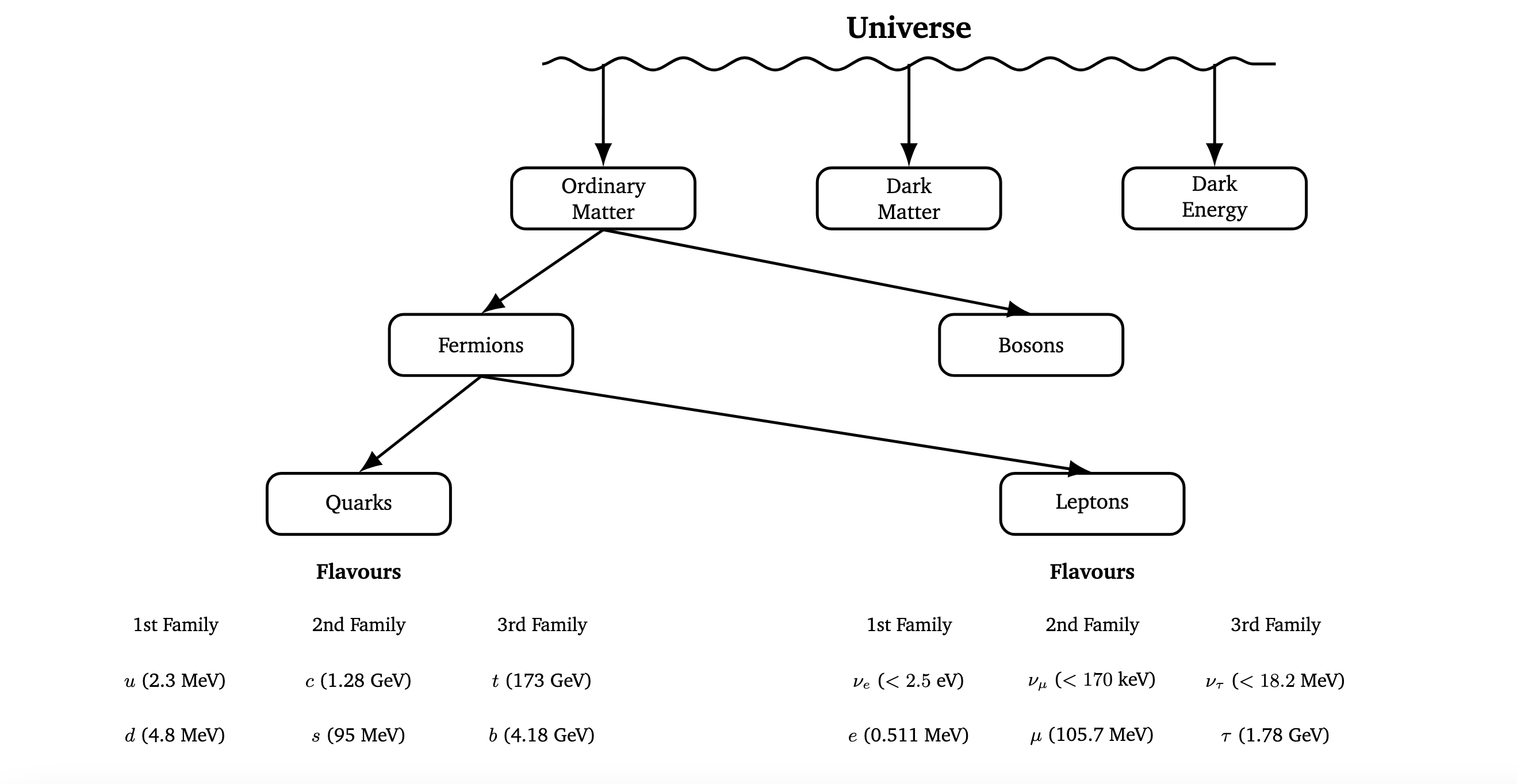}
 \caption{ The flavour problem  }
  \label{fig3}
	\end{figure}

This structure fails to explain the observed hierarchical mass spectrum of the charged fermions, as illustrated in Figure~\ref{fig3}, as well as the hierarchy in quark mixing. Furthermore, the SM lacks a mechanism to generate neutrino masses and mixing. A genuine solution to the flavour problem must not only account for the origin of neutrino masses and oscillations but also correctly predict their mass ordering and mixing structure.

We now discuss the DTC-paradigm, which provides a compelling mechanism for achieving electroweak symmetry breaking dynamically, while also reproducing the observed Higgs boson mass. Remarkably, the DTC paradigm revives conventional technicolour (TC) theories, models that were largely set aside in the early 1980s due to their tension with electroweak precision data and difficulties in generating realistic fermion masses. In this article, we highlight that the DTC approach can overcome these issues, offering a viable and consistent extension of the SM. Before introducing the details of the DTC paradigm, we briefly review the original TC idea and its major theoretical challenges.


  \section{Technicolour}
Let us consider a single fermion doublet transforming under the gauge symmetry group
$SU(3)_c \times SU(2)_L \times U(1)_Y \times SU(N_{\rm TC})$ as follows~\cite{Weinberg:1975gm,Susskind:1978ms}:
\begin{align}
T_q &\equiv 
\begin{pmatrix}
T \\
B
\end{pmatrix}_L 
: (1,2,0,N_{\rm TC}), \\[5pt]
T_R & : (1,1,+\tfrac{1}{2},N_{\rm TC}), \qquad 
B_R : (1,1,-\tfrac{1}{2},N_{\rm TC}),
\end{align}
where $T$ and $B$ carry electric charges of $+\tfrac{1}{2}$ and $-\tfrac{1}{2}$, respectively.

Moreover, if we switch off the electroweak gauge interactions, the model exhibits a global chiral symmetry given by $SU(2)_L \times SU(2)_R$. This symmetry acts on the left- and right-handed technifermions. The strong dynamics associated with $SU(N_{\rm TC})$ leads to the formation of a chiral condensate, $\langle \bar{T} T \rangle \neq 0$, which spontaneously breaks the global symmetry down to its diagonal subgroup, $SU(2)_{L+R}$. This symmetry breaking pattern is analogous to the chiral symmetry breaking in QCD, and it gives rise to three Goldstone bosons that become the longitudinal components of the $W^\pm$ and $Z$ bosons when electroweak interactions are restored.

We define the effective scalar and pseudoscalar fields associated with the technifermion bilinears as
\begin{align*}
\sigma \equiv \bar{T} T, \qquad \Pi^i \equiv i \bar{T} \tau^i \gamma_5 T,
\end{align*}
where $\tau^i$ ($i = 1, 2, 3$) are the Pauli matrices acting in isospin space. The scalar field $\sigma$ corresponds to the chiral condensate responsible for the SSB, while the $\Pi^i$ represent the associated pseudo-Nambu–Goldstone bosons (technipions). 


Now, let us turn on the electroweak interactions. In this context, the effective scalar and pseudoscalar fields can be organized into a complex scalar doublet $\varphi$ transforming under $SU(2)_L \times U(1)_Y$ as follows:
\begin{align}
\varphi = \begin{pmatrix}
\Pi_1 + i \Pi_2 \ \\
\sigma + i \Pi_3
\end{pmatrix}.
\end{align}
This composite field $\varphi$ plays the role of the Higgs doublet in the effective theory. The VEV of the scalar component,
\begin{align}
  \langle \sigma \rangle_0 = v = 246 \rm GeV, 
\end{align}
spontaneously breaks the electroweak symmetry $SU(2)_L \times U(1)_Y$ down to the electromagnetic subgroup $U(1)_{\text{EM}}$. The three pseudoscalar components $\Pi_{1,2,3}$ become the longitudinal components of the $W^\pm$ and $Z$ bosons, giving them mass via the Higgs mechanism. For more details see Ref.~\cite{Cheng:1984vwu}.

\subsection{Fermion masses}
In conventional TC models, it is typically assumed that the gauge symmetries of both the SM and the TC sector are embedded within a larger extended TC (ETC) gauge group. The masses of SM fermions are then generated through interactions mediated by the ETC gauge bosons and is given by~\cite{Hill:2002ap}:
\begin{equation}
m_f \propto \frac{\Lambda_{\rm TC}^{3}}{\Lambda_{\rm ETC}^2},
\label{eq:fermion_mass_TC}
\end{equation}
where $\Lambda_{\rm TC}$ is the TC confinement scale and $\Lambda_{\rm ETC}$ is the scale of the ETC interactions. However, flavour-changing neutral current (FCNC) constraints impose a severe lower bound on the ETC scale, typically requiring~\cite{Hill:2002ap}:
\begin{equation}
\Lambda_{\rm ETC} \gtrsim 10^6~\text{GeV}.
\end{equation}
This large separation between the TC and ETC scales leads to an extreme suppression of fermion masses, making it difficult to realistically reproduce the observed SM fermion mass spectrum.

\subsection{Higgs mass}
The lightest scalar in QCD, the $\sigma$ meson, can be estimated using the relation \cite{delbourgo1982}:
\begin{equation}
\label{sigma_mass}
m_{\sigma} \approx 2 m_{\rm dyn},
\end{equation}
where $m_{\rm dyn}$ is the non-perturbatively generated dynamical mass. The mass of the $\sigma$ meson turns out to be  $m_\sigma \approx 500$ MeV where   $m_{\rm dyn} \approx \Lambda_{\rm QCD} \approx 250$ MeV is assumed.  This prediction is consistent with the experimental observation \cite{pdg24}.

By analogy, the mass of a composite Higgs boson in a QCD-like TC theory can be estimated, and is given by \cite{Elias:1984zh}:
\begin{equation}
m_{\rm H} \approx 2 M_{\rm dyn, TC}.
\end{equation}
Even if we take the dynamical TC scale to be $M_{\rm dyn, TC} \approx 100$ GeV, roughly matching the electroweak scale, the resulting prediction for the Higgs mass is approximately 200 GeV. This is substantially higher than the experimentally observed Higgs mass of 125 GeV, as reported by the ATLAS and CMS collaborations \cite{atlas,cms}. Consequently, QCD-like TC theories encounter a significant problem, as they tend to predict a Higgs boson that is too heavy, in clear conflict with experimental value of the Higgs mass.
\subsection{Electroweak constraint }
The dynamics of TC theories are strongly constrained by electroweak precision observables, particularly the oblique parameters $S$ and $T$ \cite{Peskin:1990zt,Peskin:1991sw}. Among these, the $S$ parameter is especially sensitive to the particle content and spectrum of TC models. The current values, as reported by the Particle Data Group, are \cite{pdg24}:
\begin{equation}
S = -0.04 \pm 0.10, \quad T = 0.01 \pm 0.12.
\end{equation}
These tight bounds pose a significant challenge to QCD-like TC theories, which tend to generate large positive contributions to the $S$ parameter.

Furthermore, constraints on the vector ($V$) and axial-vector ($A$) resonances in QCD-like theories have been derived. According to Refs.~\cite{Pich:2025ywr,Pich:2013fea}, the following bounds apply at 95 $\%$ confidence level:
\begin{equation}
\label{s_bound}
M_A \geq M_V \geq 2~\text{TeV}.
\end{equation}
These limits further restrict the viable parameter space for conventional TC scenarios, motivating the search for alternative frameworks or mechanisms that can soften their contributions to the oblique parameters.

\section{Return of the technicolour}
One of the key reasons for the failure of conventional TC theories lies in the issue of hierarchy. For instance, a large hierarchy is required between the TC scale and the ETC scale in order to suppress FCNCs. However, this suppression also leads to an extreme reduction in the generated fermion masses, rendering them unrealistic.  This leads to us to itroduce the DTC paradigm.

    \subsection{Dark-technicolour paradigm}
We now turn our attention to the DTC paradigm~\cite{Abbas:2020frs}, which provides a unified framework wherein both the Standard Hierarchical Vacuum-Expectations Model (SHVM)~\cite{Abbas:2024jut,Abbas:2023dpf,Abbas:2017vws} and the Froggatt–Nielsen (FN) mechanism \cite{Froggatt:1978nt} based on the  $\mathcal{Z}_{\rm N} \times \mathcal{Z}_{\rm M}$ flavour symmetry~\cite{Abbas:2018lga,Abbas:2022zfb,Abbas:2023ion,Abbas:2024dfh},  naturally emerge at low energies~\cite{Abbas:2025ser}. Originally proposed in~\cite{Abbas:2020frs}, the DTC paradigm is based on the following key assumptions:
\begin{enumerate}
    \item  
    The fundamental symmetry group is given by
    \begin{equation}
    \mathcal{G} \equiv \text{SU}(\text{N}_{\text{TC}}) \times \text{SU}(\text{N}_{\text{DTC}}) \times \text{SU}(\text{N}_{\text{D}}),
    \end{equation}
    which consists of QCD-like gauge symmetries, and $\rm D$ represents the dark-QCD (DQCD) dynamics.
    
    \item
    All fermion masses and mixings, including those of neutrinos, are generated dynamically via multi-fermion condensates. At low energies, these condensates behave as hierarchical VEVs, effectively reproducing the SHVM and addressing the flavour structure of the Standard Model.
    
    \item
    The DTC framework incorporates the EMAC hypothesis, which solves the flavour problem of the SM.
\end{enumerate}
The TC sector consists of TC fermions transforming under the full gauge group $\mathrm{SU}(3)_c \times \mathrm{SU}(2)_L \times \mathrm{U}(1)_Y \times \mathcal{G}$ as follows:
\begin{eqnarray}
T  &\equiv&  \begin{pmatrix}
T  \\
B
\end{pmatrix}_L :(1,2,0,\rm{N}_{\rm TC},1,1), ~
T_{R}^i : (1,1,1,\text{N}_{\rm{TC}},1,1), B_{R}^i : (1,1,-1,\rm{N}_{\rm TC},1,1), 
\end{eqnarray}
where $i = 1,2,3,\dots$, and the electric charges are $Q_T = +\frac{1}{2}$ and $Q_B = -\frac{1}{2}$. For collider studies, we focus on a minimal TC sector with a single fermion doublet, i.e., two flavours.

The DTC-sector, transforming under $\mathcal{G} $, includes the following Dirac fermions:    
\begin{eqnarray}
 D^i &\equiv& C_{L,R}^i  : (1,1, 1,1,\rm N_{\rm DTC},1),~S_{L,R}^i  : (1,1,-1,1,\rm N_{\rm DTC},1), 
\end{eqnarray}
where $i=1,2,3,\dots$, and the electric charges are $Q_C = +\frac{1}{2}$ and $Q_S = -\frac{1}{2}$.  

Lastly, the DQCD sector, governed by $\mathcal{G} $, features fermions transforming as:
\begin{eqnarray}
F_{L,R} &\equiv &U_{L,R}^i :  (3,1,\dfrac{4}{3},1,1,\rm N_{\rm D}),
D_{L,R}^{i} :   (3,1,-\dfrac{2}{3},1,1,\rm N_{\rm D}),  \\ \nonumber 
&& N_{L,R}^i :   (1,1,0,1,1,\rm N_{\rm D}), ~E_{L,R}^{i} :   (1,1,-2,1,1,\rm N_{\rm D}),
\end{eqnarray}
with $i = 1,2,3,\dots$. These fields are vector-like under the Standard Model gauge symmetries.

We assume that the TC fermions, left-handed SM fermions, and $F_R$ fermions are unified under an ETC symmetry, while an Extended DTC (EDTC) symmetry similarly unifies the DTC fermions, right-handed SM fermions, and $F_L$ fermions. The $\mathrm{SU}(N_{\mathrm{D}})$ gauge group, corresponding to the DQCD sector, serves as a bridge between the TC and DTC dynamics. This connection naturally suppresses mixing between the two sectors by a factor of $1/\Lambda$, where $\Lambda$ is the confinement scale of DQCD.

In the DTC framework, the full symmetry group $\mathcal{G}$ gives rise to three global axial symmetries that are classically conserved but quantum mechanically anomalous: $\rm U(1)_A^{TC}$, $\rm U(1)_A^{DTC}$, and $\rm U(1)_A^{D}$. These $\rm U(1)_A$ symmetries can be broken by non-perturbative effects such as instantons, which are a characteristic feature of strongly coupled gauge theories. The instanton background leads to the formation of effective multi-fermion interactions, specifically, $2K$-fermion operators that develop nonzero VEVs~\cite{Harari:1981bs}. As a result, the original continuous symmetry is broken down to a discrete subgroup:
\begin{equation}
\rm U(1)_A \rightarrow \mathcal{Z}_{2K},
\end{equation}
where $K$ is the number of massless fermion flavours transforming under the fundamental representation of the corresponding $\rm SU(N)$ gauge group. This residual discrete symmetry plays an important role in determining the low-energy structure and the spectrum of the theory, as first discussed in \cite{Harari:1981bs}.

Consequently, the DTC framework generically leads to a residual discrete flavour symmetry of the form:
\begin{align}
 \mathcal{Z}_N \times \mathcal{Z}_M \times \mathcal{Z}_P   
\end{align}
where $\rm N = 2K_{\rm TC}$, $\rm M = 2K_{\rm DTC}$, and $\rm P = 2K_{\rm D}$. These discrete symmetries correspond to conserved axial charges modulo $2K$, as originally noted in \cite{Harari:1981bs}.

    \subsection{The extended most attractive channel hypothesis}
Aoki and Bando (AB), in a series of seminal papers \cite{Aoki:1983ae,Aoki:1983za,Aoki:1983yy}, demonstrated that multifermion states of the form $(\bar{\psi}_L \psi_R)^n$, composed of $2n$ fermions, exhibit an increasingly attractive interaction as the number of fermion pairs $n$ increases. This enhanced attraction arises from the underlying spin and chiral structure of the system and can be systematically characterized in terms of these features \cite{Aoki:1983yy}. Before turning to the experimental constraints relevant to the DTC paradigm, we briefly summarize the key principles of this framework.

As a concrete illustration, consider a two-fermion system in a non-Abelian gauge theory, as studied in \cite{Aoki:1983ae,Aoki:1983za,Aoki:1983yy}. If the interaction between the fermions is mediated by a single gauge boson exchange, the effective potential is given by:
\begin{align}
\label{pot}
V = g^2, F(i_1, i_2; i_1', i_2') , \langle \lambda^a(1), \lambda^a(2) \rangle,
\end{align}
where $\lambda^a(n)$ are the generators of the $\mathrm{SU}(N)$ gauge group acting on the $n$th fermion, and $g$ denotes the gauge coupling constant. The indices $i_n^{(\prime)}$ encapsulate all non-colour degrees of freedom, such as momentum, spin, and chirality.

In the ``most attractive channel" (MAC) hypothesis \cite{Raby:1979my}, the interaction factor $F$  is assumed to be universal for fermions residing in different representations of the gauge group $\rm SU(N)$. Within this framework, condensation is allowed only in the $\psi_L \psi_L$ or $\bar{\psi}_L \psi_R$ channels. Consequently, the factor $F$ plays a relatively trivial role in determining the dynamics of condensation under the MAC assumption.

In the EMAC hypothesis~\cite{Aoki:1983ae,Aoki:1983za,Aoki:1983yy}, the interaction factor \( F \) acquires a nontrivial dependence on the chirality of the participating fermions, specifically through the number and configuration of fermions involved in a given chiral condensate. AB argue that, due to the attractive nature of the potential in Eq.~\eqref{pot}, a chiral condensate forms at a dynamical scale \( \mu \), which is determined by the condition:
\begin{align}
V \left(g^2(\mu^2)\right) \sim 1,
\end{align}
where the running gauge coupling is given by:
\begin{align}
g^2(\mu^2) = \frac{1}{\beta_0 \log(\mu^2/\Lambda^2)}.
\end{align}
Solving this condition yields the condensation scale:
\begin{align}
\mu^2 \sim \Lambda^2 \exp\left( -\frac{F \langle \lambda \lambda \rangle}{\beta_0} \right).
\end{align}

This expression reveals that even small differences in chiral structure, encoded in the factor \( F \), are exponentially magnified in the formation of condensates. As a result, channels with more favorable chiral configurations dominate dynamically. However, since the exact structure of \( F \) is not fully known, a complete quantitative description of the hierarchy among competing chiral channels remains an open question.

In the case of a two-fermion system, the EMAC hypothesis identifies $\bar{\psi}_R \psi_L$ and $\psi_L \psi_L$ as the most attractive channels in the spin-zero sector, consistent with the traditional MAC hypothesis. For spin-one, colour-singlet states, the configuration with highest chiral attraction is $\bar{\psi}_L \psi_L$. Further technical details and derivations can be found in Refs.~\cite{Aoki:1983ae,Aoki:1983za,Aoki:1983yy}.

In general, an \( n \)-body colour-singlet, spin-zero multifermion condensate of the form \( (\bar{\psi}_R \psi_L)^{n/2} \), possessing maximal chirality for even \( n \), can be characterized by its average energy as follows~\cite{Aoki:1983ae,Aoki:1983za,Aoki:1983yy}:
\begin{align}
\bar{E}(n) = \frac{1}{n} E(\bar{\psi}_R^{n/2} \psi_L^{n/2}) \lesssim V_E^{LL} \frac{N^2 - 1}{N} - V_M^{LL} \frac{N - 1}{N}(n + 3N + 1),
\end{align}
where \( V_E^{LL} \) and \( V_M^{LL} \) denote the electric and magnetic components of the two-body Hamiltonian, respectively.

The above expression shows that \( \bar{E}(n) \) decreases linearly with \( n \), indicating that multifermion systems become increasingly attractive as the number of fermions grows. Consequently, a natural hierarchy emerges among the multifermion chiral condensates, following the pattern:
\begin{align}
\label{mult_cond}
\langle \bar{\psi}_R \psi_L \rangle \ll \langle \bar{\psi}_R \bar{\psi}_R \psi_L \psi_L \rangle \ll \langle \bar{\psi}_R \bar{\psi}_R \bar{\psi}_R \psi_L \psi_L \psi_L \rangle \ll \cdots.
\end{align}
This series terminates at \( n_{\text{max}} \), which is bounded by the number of distinct fermion species available in the theory~\cite{Aoki:1983ae,Aoki:1983za,Aoki:1983yy}.

The interaction factor \( F \) exhibits a proportional dependence on the net chirality \( \Delta \chi \) of the multifermion operator:
\begin{align}
F \propto \Delta \chi,
\end{align}
where \( \Delta \chi \) denotes the total chirality carried by the condensate. This leads to a parametrization of the hierarchical structure of chiral multifermion condensates as~\cite{Aoki:1983yy}:
\begin{align}
\label{VEV_h}
\langle ( \bar{\psi}_R \psi_L )^n \rangle \sim \left( \Lambda \exp(k \Delta \chi) \right)^{3n},
\end{align}
with \( k \) being a constant characterizing the chirality dependence, and \( \Lambda \) representing the confinement scale of the non-Abelian gauge dynamics.

    \subsection{Effective low energy limits of the DTC paradigm}
\begin{figure}[H]
\centering
\includegraphics[width=0.6\linewidth]{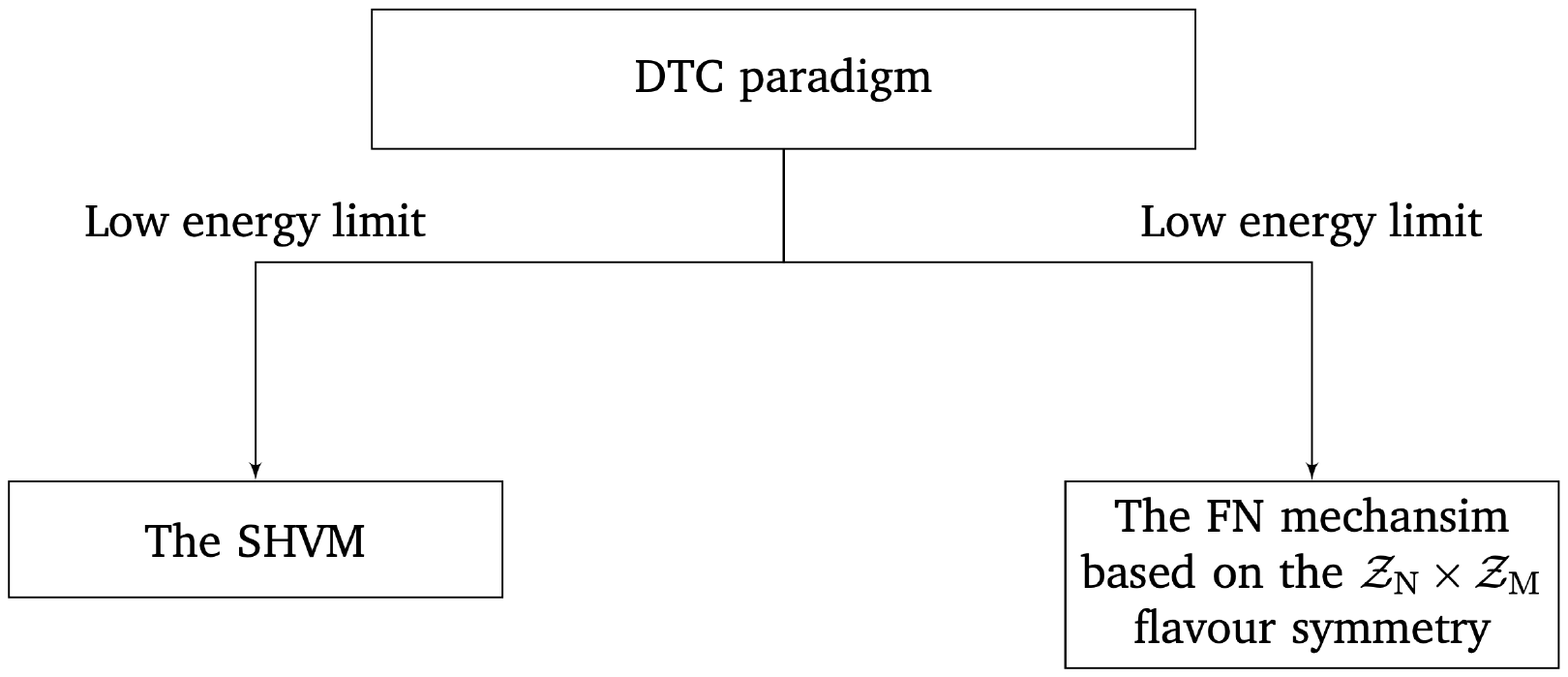}
\caption{At low energies, the DTC paradigm may effectively reduce to either the SHVM or the Froggatt–Nielsen mechanism, depending on the prevailing symmetry-breaking pattern.}
\label{fig:effective_limits}	
\end{figure}

   At low energies, the DTC paradigm can be effectively mapped onto either the SHVM or the FN mechanism based on the  discrete symmetry \( \mathcal{Z}_{\rm N} \times \mathcal{Z}_{\rm M} \), as illustrated in Figure~\ref{fig:effective_limits}. In this section, we examine how these low-energy effective descriptions emerge from the fundamental dynamics governing the DTC framework.

The chiral condensates in a QCD-like theory, following the hierarchical structure described in Eq.~\eqref{VEV_h}, are given by~\cite{Miransky:1994vk}:
\begin{align}
\label{chi_con}
\langle \bar{T} T \rangle_{\Lambda_{\rm ETC}} &\approx - \frac{N_{\rm TC}}{4\pi^2} \left[ \Lambda_{\rm TC} \exp(k_{\rm TC} \Delta \chi_{\rm TC}) \right]^3, \nonumber \\
\langle \bar{D} D \rangle_{\Lambda_{\rm EDTC}} &\approx - \frac{N_{\rm DTC}}{4\pi^2} \left[ \Lambda_{\rm DTC} \exp(k_{\rm DTC} \Delta \chi_{\rm DTC}) \right]^3, \nonumber \\
\langle \bar{F} F \rangle_{\Lambda_{\rm GUT}} &\approx - \frac{N_{\rm D}}{4\pi^2} \left[ \Lambda \exp(k_{\rm D} \Delta \chi_{\rm D}) \right]^3.
\end{align}

\begin{figure}[H]
	\centering
 \includegraphics[width=0.8\linewidth]{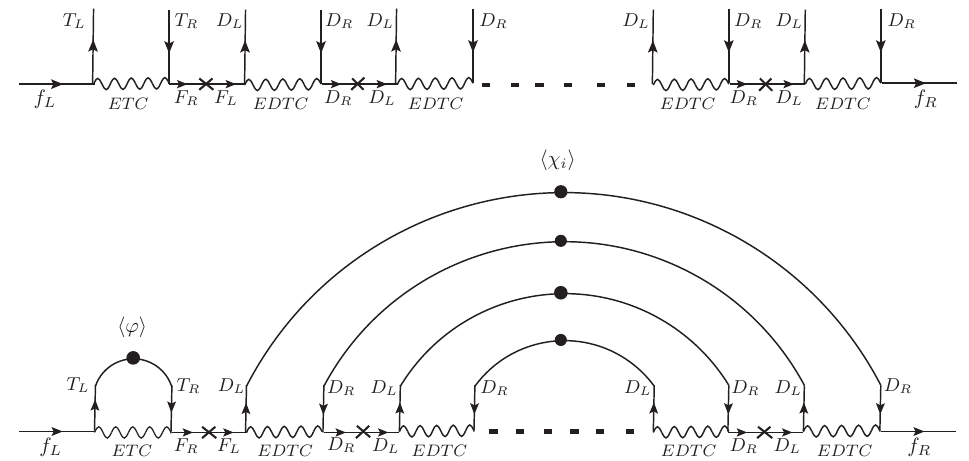}
    \caption{The Feynman diagrams for the masses of  charged fermions in the  DTC paradigm.  The top part shows the generic interactions of the SM, TC, DQCD and DTC fermions.  In the lower part of figure, the formations of the TC  chiral condensates, $\langle  \varphi \rangle$ (circular blob), a   generic   multifermion  chiral condensates  $\langle  \chi_r \rangle$ (collection of circular blobs),    and the resulting mass of the SM charged fermion is depicted. }
 \label{fig_shvm}	
 \end{figure}

Figure~\ref{fig_shvm} schematically illustrates the interactions responsible for charged fermion mass generation. The upper panel shows the generic interaction vertices among SM, TC, DQCD, and DTC fermions. The lower panel depicts the resulting mass-generation mechanism: the chiral condensate $\langle \varphi \rangle$ acts analogously to the Higgs VEV, while the DTC-induced multifermion condensates are denoted by $\langle \chi_r \rangle$.

Using Eq.~\eqref{chi_con}, the mass matrices for the up-type quarks, down-type quarks, and charged leptons can be expressed as:
\begin{align}
\label{TC_masses2}
\M_{\U,\D,\ell} & = & y_{ij}^f {{\rm N}_{\rm D}~ {\rm N}_{\rm DTC}^{(n_i-2)/2}  \dfrac{\rm N_{\rm TC}}{4 \pi^2 }  \frac{\Lambda_{\rm TC}^{3}}{\Lambda_{\rm ETC}^2}} \exp(6 k_{\rm TC} )   \dfrac{1}{\Lambda} \left[\dfrac{\rm {N_{DTC}}}{4 \pi^2 }\right]^{n_i/2} \frac{\Lambda_{\rm DTC}^{ n_i + 1}}{\Lambda_{\rm EDTC}^{n_i}} \left[\exp(6 k_{\rm DTC}) \right]^{n_i/2},
\end{align}
where we assume the TC chiral condensate is of the form \( \langle \bar{T}_R T_L \rangle \), corresponding to \( \Delta \chi_{\rm TC} = 2 \). Furthermore, we take the ETC and EDTC couplings to be \( g_{\rm ETC} = g_{\rm EDTC} \sim \mathcal{O}(1 - 4\pi) \), and assume \( k_{\rm TC} = 0 \). The quantity \( n_i = 2,4,6,\ldots,2n \) denotes the number of fermions involved in the multifermion DTC chiral condensate, which plays the role of the effective VEV \( \langle \chi_r \rangle \)~\cite{Abbas:2020frs}. The scales \( \Lambda_{\rm TC} \), \( \Lambda_{\rm DTC} \), and \( \Lambda \) correspond to the confinement scales of the TC, DTC, and DQCD sectors, respectively.

 At low energies, the theory reduces to the following effective Lagrangian:
 \bea
\label{mass2}
{\mathcal{L}} &=& \dfrac{1}{\Lambda }\Bigl[  y_{ij}^u  \bar{\psi}_{L_i}^{q}  \tilde{\varphi} \psi_{R_i}^{u}   \chi _r +     
   y_{ij}^d  \bar{\psi}_{L_i}^{q}   \varphi \psi_{R_i}^{d}  \chi _{r}   +   y_{ij}^\ell  \bar{\psi}_{L_i}^{\ell}   \varphi \psi_{R_i}^{\ell}  \chi _{r} \Bigr]  
+  {\rm H.c.},
\eea
$i$ and $j$   show family indices,  $ \psi_{L}^q,  \psi_{L}^\ell  $ are  the  quark and leptonic doublets,  $ \psi_{R}^u,  \psi_{R}^d, \psi_{R}^\ell$ denote the right-handed up,  down-type  quarks and  leptons,  $\varphi$ and $ \tilde{\varphi}= -i \sigma_2 \varphi^* $  are the SM Higgs field, and its conjugate, where  $\sigma_2$ denote  the second Pauli matrix.  

Moreover, the gauge-singlet scalar fields $\chi_r $  transform trivially under the Standard Model gauge group  $ \rm SU(3)_c \times SU(2)_L \times U(1)_Y$ as:
\begin{eqnarray}
\chi_r :(1,1,0),
 \end{eqnarray} 
where $r=1-6$.

The Lagrangian in Eq. \ref{mass2}, in fact, the effective Lagrangian of the SHVM, obtained by imposing a discrete flavour symmetry of the form 
\begin{align}
 \mathcal{Z}_{\rm N} \times \mathcal{Z}_{\rm M} \times \mathcal{Z}_{\rm P},   
\end{align}
as discussed in Refs.~\cite{Abbas:2017vws,Abbas:2020frs,Abbas:2023dpf,Abbas:2024jut}, and given by,
\bea
\label{mass22}
{\mathcal{L}_{f}} &=& \dfrac{1}{\Lambda }\Bigl[  y_{11}^u  \bar{\psi}_{L_1}^{q}  \tilde{\varphi} \psi_{R_1}^{u}   \chi _1 +  y_{13}^u  \bar{\psi}_{L_1}^{q}  \tilde{\varphi} \psi_{R_3}^{u}   \chi _5  +  y_{22}^u  \bar{\psi}_{L_2}^{q}  \tilde{\varphi} \psi_{R_2}^{u}   \chi_2  +  y_{23}^u  \bar{\psi}_{L_2}^{q}  \tilde{\varphi} \psi_{R_3}^{u}   \chi_2^\dagger  
  +  y_{33}^u  \bar{\psi}_{L_3}^{q}  \tilde{\varphi} \psi_{R_3}^{u}   \chi_3 \\ \nonumber
  && +   y_{11}^d  \bar{\psi}_{L_1}^{q}   \varphi \psi_{R_1}^{d}  \chi_{1} +     
   y_{12}^d  \bar{\psi}_{L_1}^{q}   \varphi \psi_{R_2}^{d}  \chi_{4}           
    + y_{22}^d  \bar{\psi}_{L_2}^{q}   \varphi \psi_{R_2}^{d}  \chi_{5}  + y_{33}^d  \bar{\psi}_{L_3}^{q}   \varphi \psi_{R_3}^{d}  \chi_{6}  \\ \nonumber 
     && +   y_{11}^\ell  \bar{\psi}_{L_1}^{\ell}   \varphi \psi_{R_1}^{\ell}  \chi _{1}  +   y_{12}^\ell  \bar{\psi}_{L_1}^{\ell}   \varphi \psi_{R_2}^{\ell}  \chi _{4}  +   y_{13}^\ell  \bar{\psi}_{L_1}^{\ell}   \varphi \psi_{R_3}^{\ell}  \chi _{5}  +   y_{22}^\ell  \bar{\psi}_{L_2}^{\ell}   \varphi \psi_{R_2}^{\ell}  \chi _{5}  +   y_{23}^\ell  \bar{\psi}_{L_2}^{\ell}   \varphi \psi_{R_3}^{\ell}  \chi _{2}^\dagger  \\ \nonumber
&& +   y_{33}^\ell  \bar{\psi}_{L_3}^{\ell}   \varphi \psi_{R_3}^{\ell}  \chi _{2} +  {\rm H.c.} \Bigr].
\eea

The gauge singlet fields $\chi_r$ acquire the VEVs   $\langle \chi_r \rangle$ after the SSB of the $\mathcal{Z}_{\rm N} \times \mathcal{Z}_{\rm M} \times \mathcal{Z}_{\rm P}$ flavour symmetry.  The mass matrices of charged fermions are now given by \cite{Abbas:2024jut},
\begin{align}
\label{mUD1}
\M_\U & =   \dfrac{ v }{\sqrt{2}} 
\begin{pmatrix}
y_{11}^u  \epsilon_1 &  0  & y_{13}^u  \epsilon_{5}    \\
0    & y_{22}^u \epsilon_{2}  &  y_{23}^u  \epsilon_{2}   \\
0   &  0    &  y_{33}^u  \epsilon_{3} 
\end{pmatrix},  
\M_\D = \dfrac{ v }{\sqrt{2}} 
 \begin{pmatrix}
  y_{11}^d \epsilon_{1} &    y_{12}^d \epsilon_{4} &  0 \\
0 &     y_{22}^d \epsilon_{5} &  0\\
  0 &    0  &   y_{33}^d \epsilon_{6}\\
\end{pmatrix},
\M_\ell =\dfrac{ v }{\sqrt{2}} 
  \begin{pmatrix}
  y_{11}^\ell \epsilon_1 &    y_{12}^\ell \epsilon_{4}  &   y_{13}^\ell \epsilon_{5} \\
 0 &    y_{22}^\ell \epsilon_{5} &   y_{23}^\ell \epsilon_{2}\\
   0  &    0  &   y_{33}^\ell \epsilon_{2} \\
\end{pmatrix},
\end{align} 
where $\epsilon_r = \dfrac{\langle \chi _{r} \rangle }{\Lambda}$ and  $\epsilon_r<1$.

The masses of charged fermions  can be written as,
\begin{eqnarray}
\label{mass1a}
m_t  &=& \ \left|y^u_{33} \right| \epsilon_{3} v/\sqrt{2}, ~
m_c  = \   \left|y^u_{22} \epsilon_{2} \right|  v /\sqrt{2} ,~
m_u  =  |y_{11}^u  |\,  \epsilon_1 v /\sqrt{2},\nonumber \\
m_b  &\approx& \ |y^d_{33}| \epsilon_{6} v/\sqrt{2}, 
m_s  \approx \   \left|y^d_{22}  \right| \epsilon_{5} v /\sqrt{2},
m_d  \approx  \left|y_{11}^d    \right|\,  \epsilon_{1} v /\sqrt{2},\nonumber \\
m_\tau  &\approx& \ |y^\ell_{33}| \epsilon_{2} v/\sqrt{2}, ~
m_\mu  \approx \   |y^\ell_{22} | \epsilon_{5} v /\sqrt{2} ,~
m_e  =  |y_{11}^\ell   |\,  \epsilon_1 v /\sqrt{2}.
\end {eqnarray}

\begin{figure}[h]
	\centering
 \includegraphics[width=0.8\linewidth]{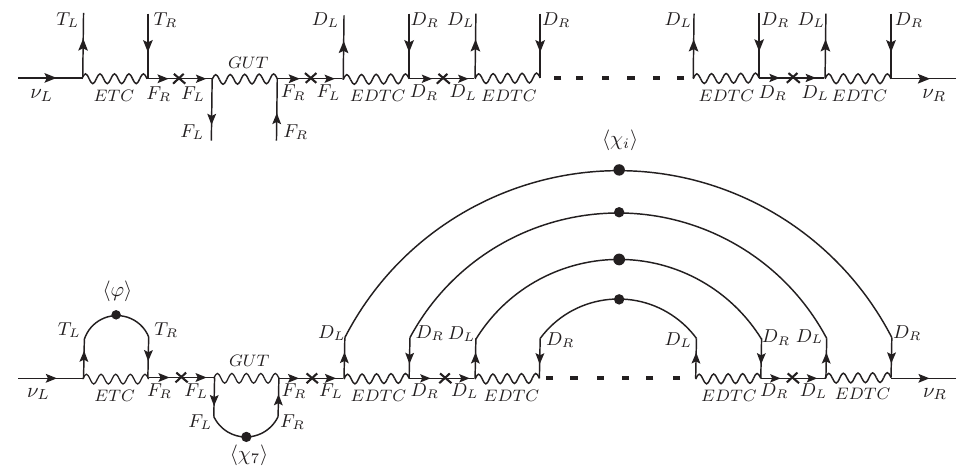}
    \caption{Feynman diagrams for neutrino mass generation in the DTC paradigm. The top panel shows interactions among SM, TC, DQCD, and DTC sectors via ETC, EDTC, and GUT bosons. The bottom panel illustrates the effective diagram after condensate formation. }
 \label{fig_nu}	
 \end{figure}

The quark mixing angles are given by,
\begin{eqnarray}
\sin \theta_{12}  & \simeq&   \left|\frac{ y_{12}^d}{ y_{22}^d} \right| { \epsilon_{4} \over \epsilon_{5}}, ~ 
\sin \theta_{23}  \simeq   \left|\frac{ y_{23}^u}{ y_{33}^u} \right| { \epsilon_{2} \over \epsilon_{3}}, 
\sin \theta_{13}  \simeq    \left|\frac{ y_{13}^u}{ y_{33}^u} \right| { \epsilon_{5} \over \epsilon_{3}}.
\end{eqnarray} 
To generate neutrino masses, we assume that the ETC and EDTC symmetries are unified within a grand unified theory (GUT). This unification gives rise to effective dimension-6 operators, as presented in equation~\ref{mass_N}, from which the neutrino mass terms emerge. The corresponding interactions are depicted in the upper part of Figure~\ref{fig_nu}, where GUT gauge bosons mediate interactions between the \(F_L\) and \(F_R\) fermions. The chiral condensate \(\langle \bar{F}_L F_R \rangle\), illustrated as a circular blob, plays a role analogous to a scalar VEV, denoted by \(\langle \chi_7 \rangle\). The lower part of Figure~\ref{fig_nu} schematically represents the effective formation of the neutrino mass term.

 The neutrino mass matrix  is  recovered as,
\bea
\label{TC_nmassesN}
\M_{\N}  =  y_{ij}^\nu {\rm N_{\rm D}}^2 {\rm N}_{\rm DTC}^{(n_i-2)/2}\dfrac{{\rm N_{\rm TC}}}{4 \pi^2 }  \frac{\Lambda_{\rm TC}^{3}}{\Lambda_{\rm ETC}^2} \exp(6 k_{\rm TC} )   \dfrac{1}{\Lambda} \left[\dfrac{{\rm N_{\rm DTC}}}{4 \pi^2 }\right]^{n_i/2} \frac{\Lambda_{\rm DTC}^{n_i + 1}}{\Lambda_{\rm EDTC}^{n_i}} \left[\exp(6 k_{\rm DTC}) \right]^{n_i/2} \dfrac{1}{\Lambda} \dfrac{{\rm N_{D}}}{4 \pi^2 } \frac{\Lambda^{3}}{\Lambda_{\rm GUT}^{2}} \exp(6 k_{\rm D}),
\eea

The interactions illustrated in Figure~\ref{fig_nu} lead, at low energies, to the following effective Lagrangian:
\begin{eqnarray}
\label{mass_N}
-{\mathcal{L}}_{\rm Yukawa}^{\nu} &=& y_{ij}^\nu \, \bar{ \psi}_{L_i}^\ell \, \tilde{\varphi} \, \nu_{f_R} \left[  \dfrac{ \chi_r \chi_7 \, (\text{or}~ \chi_r \chi_7^\dagger)}{\Lambda^2} \right] +  {\rm H.c.}.  
\end{eqnarray}
The above Lagrangian, after imposing the $ \mathcal{Z}_{\rm N} \times \mathcal{Z}_{\rm M} \times \mathcal{Z}_{\rm P} $ flavour symmetry, becomes~\cite{Abbas:2017vws,Abbas:2020frs,Abbas:2023dpf,Abbas:2024jut},
\bea
\label{mass2c}
{\mathcal{L}_{\nu}} &=& \dfrac{1}{\Lambda^2 }\Bigl[   y_{11}^\nu  \bar{\psi}_{L_1}^{\ell}  \tilde{\varphi} \psi_{R_1}^{\nu}   \chi _1^\dagger \chi_7^\dagger +  y_{12}^\nu  \bar{\psi}_{L_1}^{\ell}  \tilde{\varphi} \psi_{R_2}^{\nu}   \chi _4^\dagger \chi_7 +  y_{13}^\nu  \bar{\psi}_{L_1}^{\ell}  \tilde{\varphi} \psi_{R_3}^{\nu}   \chi _4^\dagger \chi_7 +  y_{22}^\nu  \bar{\psi}_{L_2}^{\ell}  \tilde{\varphi} \psi_{R_2}^{\nu}   \chi _4 \chi_7 \\  \nonumber
&& +  y_{23}^\nu  \bar{\psi}_{L_2}^{\ell}  \tilde{\varphi} \psi_{R_3}^{\nu}   \chi _4 \chi_7 +  y_{32}^\nu  \bar{\psi}_{L_3}^{\ell}  \tilde{\varphi} \psi_{R_2}^{\nu}   \chi_5 \chi_7 +  y_{33}^\nu  \bar{\psi}_{L_3}^{\ell}  \tilde{\varphi} \psi_{R_3}^{\nu}   \chi_5 \chi_7 +  {\rm H.c.} \Bigr],
\eea
and Eq.~\ref{TC_nmassesN} reduces to the following Dirac neutrino mass matrix:
\begin{equation}
\label{NM}
\M_{\D} = \dfrac{v}{\sqrt{2}}  
\begin{pmatrix}
y_{11}^\nu   \epsilon_1 \epsilon_{7}   &  y_{12}^\nu   \epsilon_{4} \epsilon_{7}  & y_{13}^\nu  \epsilon_{4}  \epsilon_{7} \\
0   & y_{22}^\nu  \epsilon_{4}  \epsilon_{7} &  y_{23}^\nu  \epsilon_{4}  \epsilon_{7} \\
0   &   y_{32}^\nu  \epsilon_{5}  \epsilon_{7}   &  y_{33}^\nu  \epsilon_{5}  \epsilon_{7}
\end{pmatrix},
\end{equation}
where $\epsilon_{7} = \frac{\langle \chi _{7} \rangle}{\Lambda} <1$.

Assuming all couplings are of order one, we arrive at the following remarkable predictions for the leptonic mixing angles:
\begin{eqnarray}
\sin \theta_{12}^\ell  &\simeq&  \left|- \frac{y_{12}^\nu}{y_{22}^\nu} + \frac{y_{12}^\ell \, \epsilon_4}{y_{22}^\ell \, \epsilon_5} + \frac{y_{23}^{\ell *} y_{13}^\nu \, \epsilon_4}{y_{33}^\ell y_{33}^\nu \, \epsilon_5} \right| 
\geq \left| - \frac{y_{12}^\nu}{y_{22}^\nu} \right| 
- \left| \frac{y_{12}^\ell}{y_{22}^\ell} + \frac{y_{23}^{\ell *} y_{13}^\nu}{y_{33}^\ell y_{33}^\nu} \right| \frac{\epsilon_4}{\epsilon_5} 
\approx 1 - 2 \sin \theta_{12}, \\[6pt]
\sin \theta_{23}^\ell  &\simeq&  \left| \frac{y_{23}^\ell}{y_{33}^\ell} - \frac{y_{23}^\nu \, \epsilon_4}{y_{33}^\nu \, \epsilon_5} \right| 
\geq \left| \frac{y_{23}^\ell}{y_{33}^\ell} \right| 
- \left| \frac{y_{23}^\nu}{y_{33}^\nu} \right| \frac{\epsilon_4}{\epsilon_5}
\approx 1 - \sin \theta_{12}, \\[6pt]
\sin \theta_{13}^\ell &\simeq& \left| - \frac{y_{13}^\nu \, \epsilon_4}{y_{33}^\nu \, \epsilon_5} + \frac{y_{13}^\ell \, \epsilon_5}{y_{33}^\ell \, \epsilon_2} \right| 
\geq \left| \frac{y_{13}^\nu}{y_{33}^\nu} \right| \frac{\epsilon_4}{\epsilon_5} 
- \left| \frac{y_{13}^\ell}{y_{33}^\ell} \right| \frac{\epsilon_5}{\epsilon_2} 
\approx \sin \theta_{12} - \frac{m_s}{m_c},
\end{eqnarray}
where $m_s / m_c = \epsilon_5 / \epsilon_2$.  Thus, we conclude that the leptonic mixing angles are predicted in the SHVM framework in terms of the Cabibbo angle and the mass ratio of the strange to charm quark.

This leads to highly precise predictions for the leptonic mixing angles \cite{Abbas:2023dpf}:
\begin{align}
 \sin \theta_{12}^\ell &=  0.55 \pm 0.00134, \nonumber \\
 \sin \theta_{23}^\ell &=  0.775 \pm 0.00067, \nonumber \\
 \sin \theta_{13}^\ell &\in [0.1413,\ 0.1509],
\end{align}
which may be tested in future neutrino oscillation experiments such as DUNE, Hyper-Kamiokande, and JUNO \cite{Huber:2022lpm}.

    \subsection{Reproducing the Higgs mass}
Scaling up two-flavor QCD suggests the lightest scalar singlet lies in the range 
$1.0~{\rm TeV}\lesssim M_{\rm H,TC}\lesssim 1.4~{\rm TeV}$, heavier than the observed Higgs 
mass~\cite{Foadi:2012bb}. Foadi, Frandsen, and Sannino (FFS) showed that TC dynamics can still 
yield a $125~{\rm GeV}$ Higgs once SM top-quark radiative corrections are included, which lower 
the TC Higgs mass toward the experimental value~\cite{atlas,cms}. The corrected mass is given by~\cite{Foadi:2012bb}
\begin{align}
    M_{\rm H}^2 = M_{\rm H,TC}^2 - 12\,\kappa^2 r_t^2 m_t^2,
\end{align}
with $r_t=1$ for an SM-like top Yukawa and $\kappa={\cal O}(1)$. For technifermions in the 
fundamental representation of $SU(N_{\rm TC})$, FFS showed that the above mass range is naturally 
realized with a single technidoublet. We therefore adopt $\Lambda_{\rm TC}=M_{\rm H,TC}=1~{\rm TeV}$ 
as our benchmark.

\subsection{S-parameter}
For our TC model we take $\Lambda_{\rm TC}=10^3~\mathrm{GeV}$ and 
$f=F_{\Pi_{\rm TC}}=246~\mathrm{GeV}$ (for two flavors). The corresponding upper bound is  
\begin{equation}
\frac{4\pi F_\Pi}{\sqrt{N}} \simeq 2186~\mathrm{GeV}.
\label{eq:NDA-bound}
\end{equation}
We conjecture that the $1/\sqrt{N_{\rm TC}}$ scaling becomes effective only for $N_{\rm TC}>3$, 
so that for $N_{\rm TC}=3$ the $\rho_{\rm TC}$ mass already saturates the bound in 
Eq.~\eqref{eq:NDA-bound}. In this case the dynamics follow the solid-curve scenario, with 
$N_{\rm TC}=3$ placing the theory close to regions A or B, while still consistent with the 
$S$-parameter constraint of Eq.~\eqref{s_bound}. This picture is further supported by recent 
lattice computations.  

Specifically, Ref.~\cite{Nogradi:2019iek} finds that for $N_c=3$ the ratio $M_\rho/F_\pi$ in 
the chiral limit is essentially independent of $N_f$:  
\begin{equation}
\frac{M_\rho}{F_\pi}\Bigg\rvert^{N_f=2-6}_{N_c=3} = 7.95(15).
\end{equation}
At large $N_c$, quenched lattice studies instead obtain~\cite{Bali:2013kia}  
\begin{equation}
\sqrt{\frac{N_c}{3}}\,\frac{M_\rho}{F_\pi}\Bigg\rvert_{N_c\to\infty} = 7.08(10).
\end{equation}

For $N_{\rm TC}=3$, Ref.~\cite{Bali:2013kia} further reports  
\begin{equation}
\frac{M_{\rho_{\rm TC}}}{\sqrt{\sigma}}=1.749(26),
\end{equation}
with the string tension related to the decay constant by  
\begin{equation}
\sqrt{\frac{3}{N_{\rm TC}}}\,\frac{F_{\Pi_{\rm TC}}}{\sqrt{\sigma}}=0.2174(30).
\end{equation}
Applying these relations with $F_{\Pi_{\rm TC}}=246~\mathrm{GeV}$ yields  
\begin{equation}\label{eq:mass_rho}
M_{\rho_{\rm TC}}=1980~\mathrm{GeV},
\end{equation}
which is consistent with the bound in Eq.~\eqref{s_bound}. A simple scaling relation 
provides an independent check~\cite{Tandean:1995ci}:  
\begin{equation}
M_{\rho_{\rm TC}}=\frac{F_{\Pi_{\rm TC}}}{f_\pi}\sqrt{\frac{3}{N_{\rm TC}}}\,m_\rho
=2007~\mathrm{GeV},
\end{equation}
again in good agreement with Eq.~\eqref{eq:mass_rho}.

\subsection{Dark-matter}
The DTC paradigm gives rise to novel classes of dark matter in its low-energy effective limits.  At lower energies, the SHVM limit allows for \textit{neutrinic dark matter}~\cite{Abbas:2024jut}, while the FN mechanism gives rise to \textit{flavonic dark matter}~\cite{Abbas:2023ion}.

    \section{Summary}
    The DTC paradigm extends conventional TC dynamics to address both electroweak symmetry breaking and the problem of the  SM flavour structure. It operates via an EMAC framework, predicting a hierarchy of chiral multifermion condensates. At low energies, DTC reduces effectively to either the FN mechanism or the SHVM, depending on the residual discrete symmetry. The paradigm introduces new classes of dark matter candidates and offers collider testable signatures~\cite{Abbas:2025ser}. The hierarchical mass structure of SM fermions emerges from the interplay of TC, dark TC, and DQCD dynamics, all embedded in an extended unification scheme through ETC and EDTC interactions.  

\begin{figure}[H]
	\centering
 \begin{subfigure}[]{0.266\linewidth}
 \includegraphics[width=\linewidth]{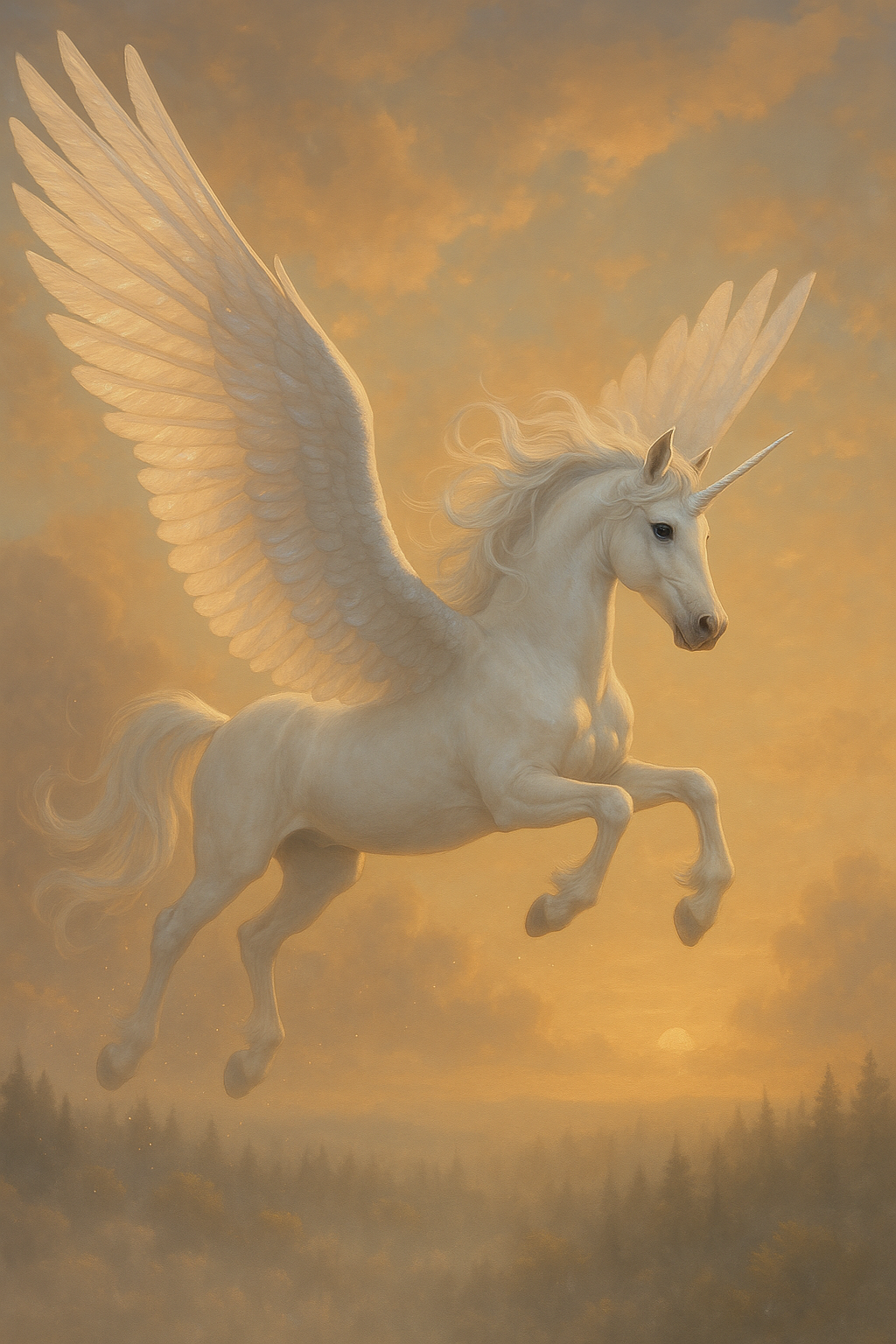}
 \caption{SUBI}
         \label{figh1a}
 \end{subfigure} 
 \begin{subfigure}[]{0.4\linewidth}
    \includegraphics[width=\linewidth]{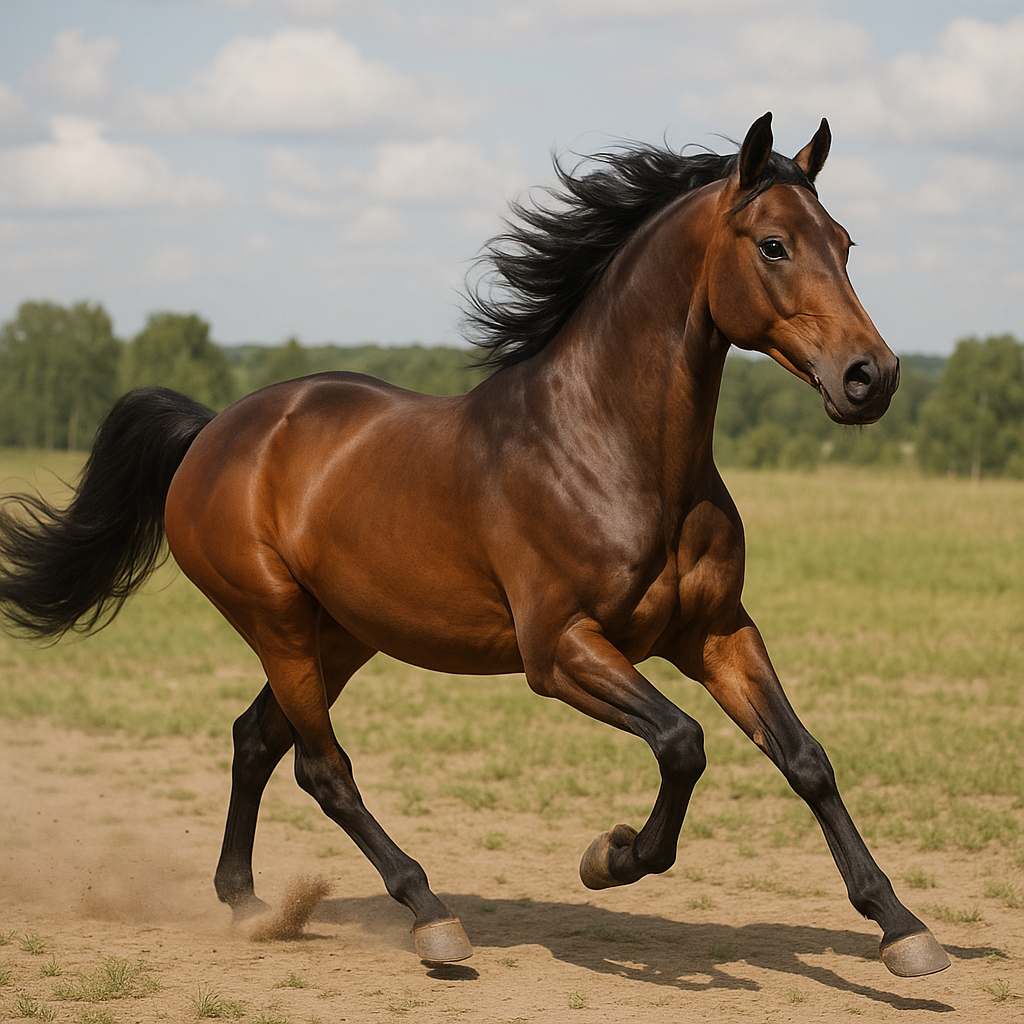}
    \caption{Reality: SWEETI}
         \label{figh1b}	
\end{subfigure}
 \caption{    }
  \label{fig_subi_sweeti}
	\end{figure}

For several decades, much of beyond-the-SM model building has been guided by the pursuit of the SUBI-type theories, highly symmetric ultraviolet completions designed to stabilize the electroweak scale by introducing new weak-scale degrees of freedom. While theoretically compelling, this approach has so far received no experimental support, despite extensive searches at the LHC. An alternative viewpoint is that nature may instead favor a more modest but structurally robust framework, in which the electroweak scale and flavour hierarchies emerge from strong dynamics rather than ultraviolet symmetries. In this sense, a SWEETI theory need not be maximally symmetric or predictive at high energies, but should instead reproduce observed phenomena through minimal assumptions and dynamical mechanisms. The dark-technicolour paradigm developed here exemplifies this philosophy, replacing ultraviolet elegance with infrared inevitability. Perhaps, as illustrated in Figure~\ref{fig_subi_sweeti}, nature prefers a SWEETI to an imposing SUBI.

\section*{Acknowledgement}
This work was supported by the Council of Science and Technology, Government of Uttar Pradesh, India, under the project \textit{``A New Paradigm for the flavour Problem''} (Project No.~CST/D-1301), and by the Anusandhan National Research Foundation (formerly SERB), Department of Science and Technology, Government of India, through the project \textit{``Higgs Physics Within and Beyond the Standard Model''} (Project No.~CRG/2022/003237). 


\end{document}